\documentclass[preprint,showpacs,preprintnumbers,amsmath,amssymb]{revtex4}

\usepackage{graphicx}
\usepackage{psfig}
\usepackage{dcolumn}
\usepackage{bm}

\begin{document}

\title{
Does a surface spin-flop occur
in antiferromagnetically coupled multilayers?\\
Magnetic states and reorientation transitions 
in antiferromagnetic superlattices
}
\author{U.K. R\"o\ss ler}
\thanks
{Corresponding author 
}
\email{u.roessler@ifw-dresden.de}

\author{A.N.\ Bogdanov}
\altaffiliation[Permanent address: ]%
{Donetsk Institute for Physics and Technology,
340114 Donetsk, Ukraine
}
\email{bogdanov@kinetic.ac.donetsk.ua}

\affiliation{
Leibniz-Institut f{\"u}r Festk{\"o}rper- 
und Werkstoffforschung Dresden\\
Postfach 270116
D--01171 Dresden, Germany
}%

\date{\today}

\begin{abstract}
{
Equilibrium spin configurations and their stability
limits have been calculated for models of magnetic superlattices 
with a finite number of thin ferromagnetic layers 
coupled antiferromagnetically through (non-magnetic) spacers
as Fe/Cr and Co/Ru multilayers. 
Depending on values of applied magnetic field 
and unaxial anisotropy, the system assumes collinear 
(antiferromagnetic, ferromagnetic, various ``ferrimagnetic'')
phases, or spatially inhomogeneous (symmetric spin-flop phase 
and asymmetric, \textit{canted} and \textit{twisted}, phases)
via series of field induced continuous and discontinuous transitions. 
Contrary to semi-infinite systems a surface phase transition,
so-called ``surface spin-flop'', does not occur in the models 
with  a finite number of layers. 
It is shown that ``discrete jumps''
observed in some Fe/Cr superlattices and 
interpreted as ``surface spin-flop'' 
transition are first-order ``volume'' 
transitions between different canted phases.
Depending on the system several of these collinear and canted phases 
can exist as metastable states  in broad ranges 
of the magnetic fields, which may cause severe hysteresis effects. 
The results explain magnetization processes 
in recent experiments on antiferromagnetic Fe/Cr superlattices.
}
\end{abstract}


         
\maketitle

%

Antiferromagnetic coupling 
in magnetic multilayers 
mediated by spacer layers
and giant magnetoresistance 
are two related phenomena
that have created the basis 
for applications of antiferromagnetic 
superlattices as Fe/Cr, Co/Cu, or Co/Ru \cite{Grunberg87}.
Multilayer stacks with antiferromagnetic 
interlayer couplings are widely used 
in spin valves 
as \textit{synthetic antiferromagnets}, 
in various other spinelectronics devices, 
and they are considered 
as promising recording media \cite{Kim01}.
High quality multilayer stacks,
such as Co/Ru \cite{Ounadjela92}, 
Fe/Cr(211) \cite{Wang94}, or Fe/Cr(001)\cite{Temst00}, 
can be considered as ``artificial'' nanoscale antiferromagnets. 
They provide experimental models 
for the magnetic properties of confined antiferromagnets
under influence of surface-effects.
Hence, both for applications 
and from a fundamental point of view, 
such systems are of great importance
and attract much interest 
in modern nanomagnetism 
\cite{Temst00,Steadman02,Felcher02,Lauter02,Nagy02,Trallori94}.

In the last years, efforts 
based on experimental investigations 
\cite{Wang94,Temst00,Steadman02,Felcher02,Lauter02,Nagy02}, 
and theoretical studies\cite{Wang94, Trallori94} 
to understand ground states and the transitions 
under magnetic fields in such multilayers 
resulted in a controversy around 
the problem of the so-called ``surface spin-flop''.
This problem can be traced back 
to Mills' theory \cite{Mills68}
which predicted that in uniaxial antiferromagnets
spins near the surfaces rotate into the flopped state
at a field reduced by a factor of $\sqrt{2}$ 
compared to the bulk spin-flop field. 
In an increasing magnetic field 
such localized surface states spread 
into the depth of the sample \cite{Mills68}.
In Ref. \cite{Wang94}, the authors claimed 
to observe these surface states 
in Fe/Cr  superlattices and supported 
their experimental results by numerical calculations. 
Subsequent theoretical studies 
(mostly based on numerical simulations 
within simplified discretized models \cite{Mills68}) 
led to conflicting conclusions 
on the evolution of magnetic states in 
these systems \cite{Trallori94}.
Finally, recent experimental investigations obtained 
different scenarios for reorientational transitions 
in Fe/Cr \cite{Felcher02,Lauter02,Nagy02},
and Co/Ru \cite{Steadman02} multilayer systems. 

This study provides a comprehensive analysis
within the standard theory of phase transitions
to determine all (one-dimensional) 
spin configurations and their stability limits
for models of antiferromagnetic superlattices.
Our results explain the diversity of experimentally 
observed effects in different antiferromagnetic 
multilayer-systems 
\cite{Wang94,Steadman02,Temst00,Felcher02,Lauter02,Nagy02}.
It is shown that the magnetization processes 
observed in \cite{Wang94} and \cite{Felcher02} 
and interpreted as a manifestation of 
the ``surface spin-flop transitions'', 
are a succession of first-order phase transitions 
between asymmetric inhomogeneous phases. 
Such transitions occur only 
in a certain range of uniaxial anisotropy.
In the major parts of the 
\textit {magnetic field-vs.-uniaxial anisotropy} 
phase diagram the antiferromagnetic phase undergoes 
discontinuous transitions
either into an inhomogeneous spin-flop phase (low anisotropy) 
or into ferrimagnetic collinear phases (high anisotropy).

The energy of a superlattice with $N$ coupled ferromagnetic
layers  can be modelled by 
\begin{eqnarray}
W&=&\sum_{i=1}^{N-1} \left[ J_i\,\mathbf{m}_i \cdot
\mathbf{m}_{i+1}
+ \widetilde{J}_i\,\left(\mathbf{m}_i \cdot
\mathbf{m}_{i+1} \right)^2 \right]
-\mathbf{H}\cdot \sum_{i=1}^{N}\mathbf{m}_i 
\nonumber \\
& - &  \frac{1}{2}\sum_{i=1}^{N} K_i\,(\mathbf{m}_i \cdot \mathbf{n})^2 
-\sum_{i=1}^{N-1} K'_i\,
(\mathbf{m}_i \cdot \mathbf{n}) (\mathbf{m}_{i+1} \cdot \mathbf{n}) 
\,,
\label{energy0}
\end{eqnarray}
where ${{\mathbf m}_i}$ are unity vectors 
along the $i$-th layer  magnetization;
the first sum includes bilinear ($J_i$) and
biquadratic ($\widetilde{J}_i$) exchange interactions;
$K_i$ and $K'_i$ are constants of uniaxial anisotropy and
$\mathbf{H}$ is an applied magnetic field. 
Here, we neglect possible variation of
magnetic parameters within the ferromagnetic
layers (see \cite{PRL01}).
As the magnetic moments of the layers
are mesoscopically large,
temperature fluctuations do not play 
a significant role for the magnetic configurations.
Thus, we have to find the zero-temperature 
ground-states described by the energy (\ref{energy0}).
Temperature dependence enters only indirectly
via the phenomenological constants 
for interlayer exchange and anisotropies.
Moreover, we consider 
the case of antiferromagnetic systems 
with fully compensated magnetization, 
i.e systems with \textit{even} number of ferromagnetic layers.
(The noncompensated magnetization 
in superlattices with odd numbers of layers or
with unequal thickness of layers strongly determines 
their magnetic properties.  Such structures are 
related to ferrimagnetic systems.
They could be studied by similar methods as used below,
but have to be considered as separate class of systems.)

The type of antiferromagnetic superlattices 
considered in our analysis are composed of few 
tens of identical magnetic/nonmagnetic
bilayers \cite{Ounadjela92,Wang94,Temst00,Felcher02,Lauter02,Nagy02,Steadman02}. 
To simplify the discussion, 
we assume that induced interactions
in such systems maintain 
mirror symmetry about the center of the layer stack,
i. e. $J_i = J_{N-i}$, $K_i=K_{N+1-i}$ etc. 
in the energy (\ref{energy0}).
Usually demagnetization fields
confine the magnetization vectors $\mathbf{m}_i$ to
the layer plane, and their orientation 
within this plane can be described by their angles
${\theta_i}$ with the easy-axis $\mathbf{n}$.
Thus the problem of the magnetic states for the
model (\ref{energy0}) is reduced to optimization
of the function $W(\theta_1, \theta_2,...\theta_N)$.
We assume that values of the magnetic parameters
are such that the energy (\ref{energy0}) 
yields a \textit{collinear} \textit{antiferromagnetic} (AF) phase 
as ground state in zero field, i.e.  $\mathbf{m}_i$ are directed 
along the ``easy axis'' $\mathbf{n}$ 
and antiparallel in adjacent layers.
Next, we consider the evolution of states with a magnetic 
field along the easy-axis  $\mathbf{n}$. 
\begin{figure}
\includegraphics[width=8.5cm]{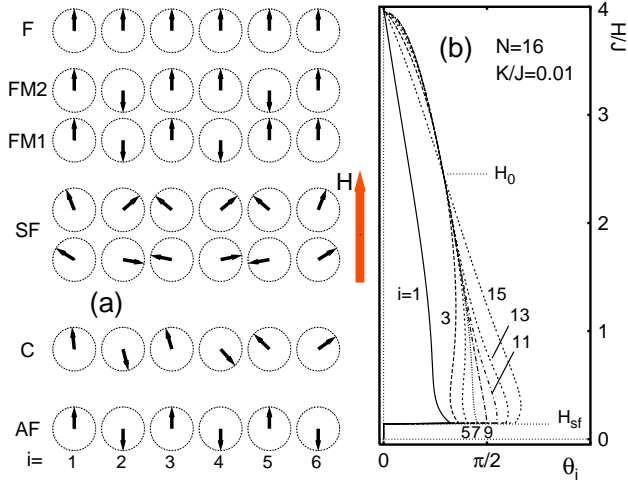}
\caption{
\label{ClocksEtc}
(a) States in antiferromagnetic superlattices (example $N=6$)
with increasing field:
F ferromagnetic; FM1/2 ferrimagnetic collinear -
such phases may be energetically degenerate,
but they own different (meta)-stability limits;
SF spin-flop states; C asymmetric canted; 
AF antiferromagnetic). 
(b) Example of evolution of state 
with field $H$ for Mills model 
in low anisotropy case:
rotation angles $\theta_{i}$ ($i$ odd) 
against easy-axis $\bf n$ $||$ field $\bf H$
(for $i$ even $\theta_{i} = - \theta_{N-i+1}$).
Phases of type C and FM1/2 may occur
only at intermediate and higher anisotropy.
}
\end{figure}

In the case of weak anisotropy
( ${ \bar{J}_i \equiv J_i-2\widetilde{J}_i \gg  K_i, K'_i}$)
the applied field stabilizes a \textit{spin-flop} (SF) phase 
with symmetric ($\theta_i=-\theta_{N-i+1}$)
deviations of $\mathbf{m}_i$ from the easy -axis (Fig. 1(a)).
Contrary to spin-flop phases in bulk antiferromagnets, 
this SF phase is spatially inhomogeneous.
At low fields the solutions for the SF phase
are given by a set of linear equations
$\bar{J}_{2j-1}(\pi -\theta_{2j-1}+\theta_{2j}) = H $, 
$\theta_{2j} - \theta_{2j+1} =0 $ ($j=1,2,...l$, $l=N/4$ for
systems with $N=4\,n$ or $l=(N+2)/4$ 
for $N=4\,n+2$, $n=0,1,\dots$).
These solutions describe small  
deviations of the magnetization vectors,
$|\theta_i - \pi/2| \ll 1$, from the directions 
perpendicular to the easy axis (Fig. 1 (a)).
Towards top and bottom layer $i=1$ or $N$ 
in the stack, the deviations increase.
For example, for $N=10$ the solutions read
$\theta_5 = \pi/2 - H/(2\bar{J}_5)$,
$\theta_4 = \theta_5 -\pi$, 
$\theta_3 = \theta_5 -H/\bar{J}_3$,
$\theta_2= \theta_3 -\pi$,
$\theta_1 = \theta_3 -H/\bar{J}_1$.
The properties of these solutions and other particular
magnetic configurations of the model (\ref{energy0})
arise essentially due 
to \textit{cut exchange bonds at the boundary layers}.
This is different from surface-induced changes 
for magnetic states of other nanoscale systems.
In ferromagnetic nanostructures, as in nanosized layers of
antiferromagnetic materials, noncollinear and/or twisted
configurations are caused by particular surface-related 
anisotropy and exchange contributions 
due to modified (relativistic) spin-orbit effects 
near surfaces (as discussed, e.g., in \cite{PRL01,InFight03}). 
The simplified variant of the energy (\ref{energy0}) 
with $J_i = J$, $K_i = K$, $\widetilde{J}_i=K'_i=0$
embodies this cutting of bonds as the only surface effect
and allows to investigate this effect  
separately from other surface-induced forces.
This model, introduced by Mills as a semi-infinite model \cite{Mills68},
was later investigated in different cases 
also for finite systems \cite{Wang94,Trallori94}.
However, in spite of rather sophisticated methods 
used in these previous studies, the magnetic properties
described by this model (called here \textit{Mills model}) 
have remained elusive.
Transitions and stability lines for the collinear phases 
can be calculated analytically, but
the main body of our results
have been obtained by numerical methods.
We could investigate 
in detail systems up to $N=20$ 
(and some aspects of larger systems) 
using a combination of methods: 
(i) search for energy minima using of the order 
1000 random starting states for a dense mesh
of points in the phase diagram;
(ii) an efficient conjugate 
gradient minimization \cite{NR} 
to solve the coupled equations
for equilibria $\{\partial W/\partial \theta_i =0\}_{i=1 \dots N}$;
(iii) calculation of stability limits 
from the evolution of the  smallest eigenvalue $e_0(H,K)$ 
of the stability matrix 
$(\partial^2 W/\partial \theta_i \partial \theta_j), i,j=1 \dots N$
under changing anisotropy constant $K$ and
the applied magnetic field. 
The basic magnetic configurations are expounded below.

\begin{figure}
\includegraphics[width=7.5cm]{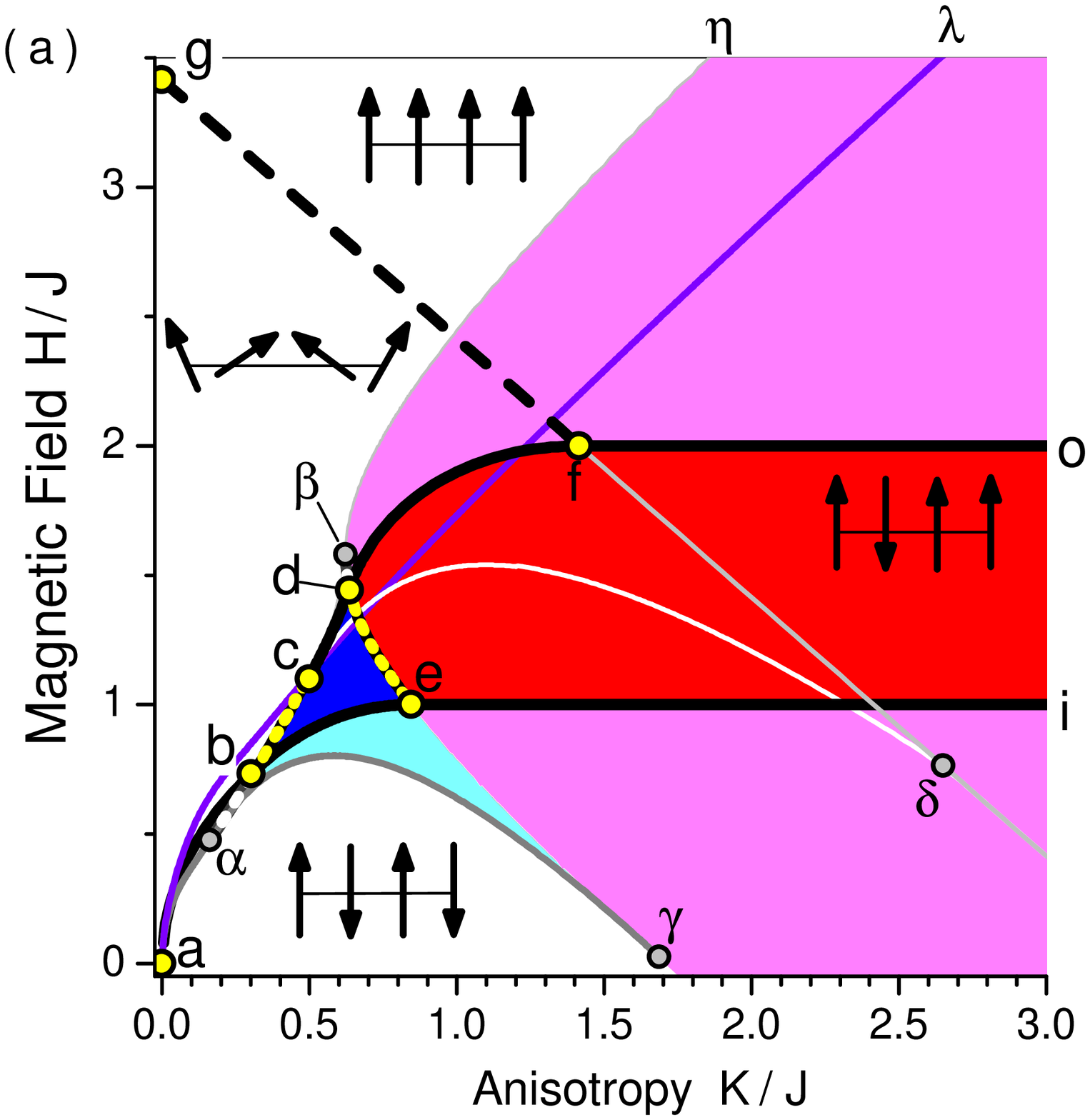}
\includegraphics[width=7.5cm]{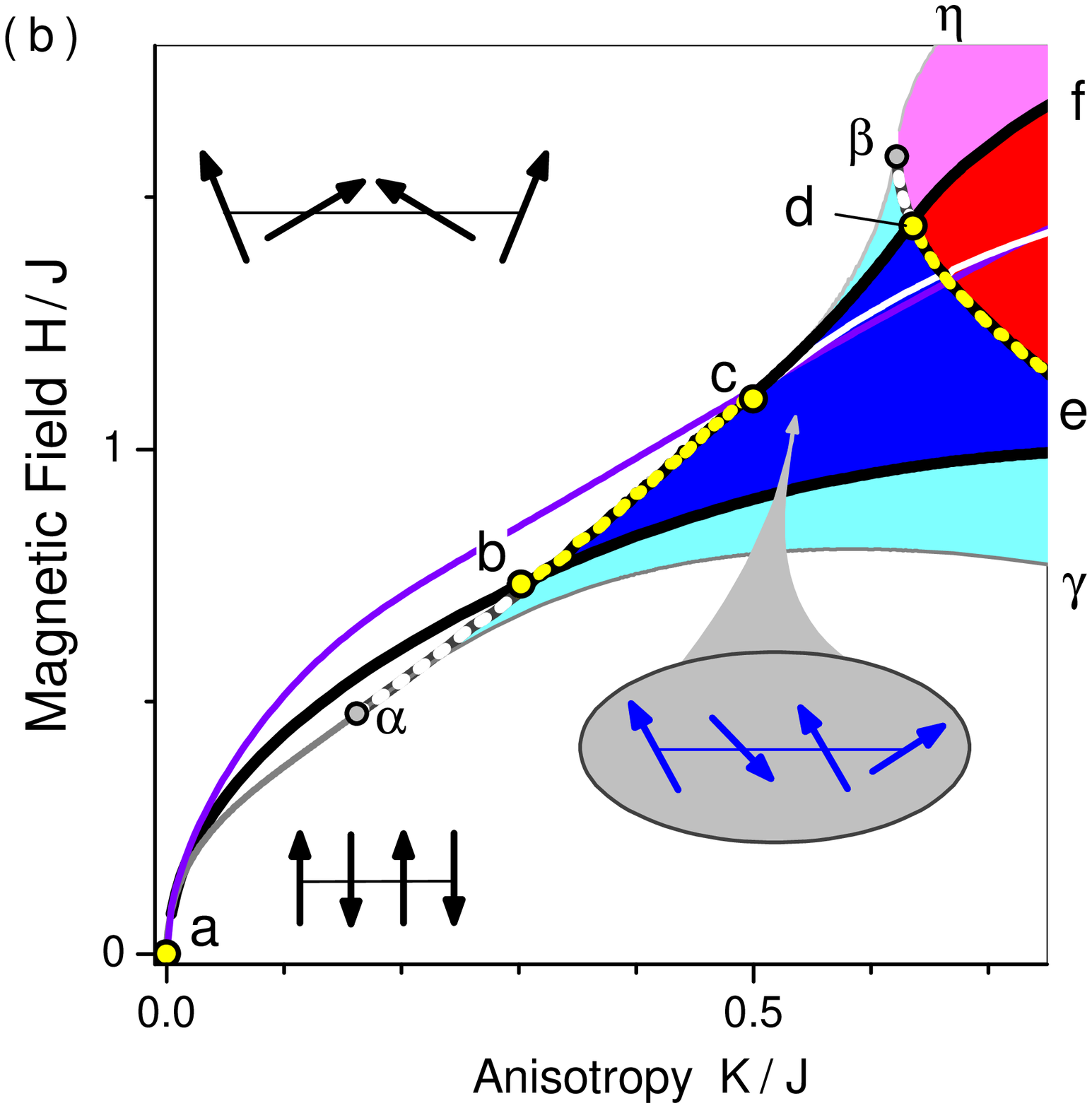}
\caption{
\label{PhaseDiagramN4}
(color) Phase-diagram for Mills model with $N=4$:
(a) overview (b) details at low anisotropy 
( in this region critical lines have been shifted 
for clarity).
Full black lines are first order transitions 
between equilibrium states;
continuous transitions are dashed and dotted. 
Equilibrium states: antiferromagnetic below $a-b-e-i$ line (AF);
(red) area $o-d-e -i$ collinear ``ferrimagnetic'' (FM);
area $a-b-d-f-g$ symmetric spin-flop phase (SF);
(blue) area $b-e-d-c$ noncollinear asymmetric (C);
above line $g-f-o $ ferromagnetic phase (FM).
Greek letters: critical points at boundaries 
of metastable states.
Metastable states corresponding to FM 
exist in the region (magenta) right of line $\eta-\beta-d-e-\gamma$ 
and for C in the two regions $\alpha-b-e-\gamma-\alpha$ 
and $c-\beta-d-c$ (light blue), respectively.
Further stability limits: for SF $a-\alpha-b-c$ 
and $c-\delta$  (white) 
$ \delta-f-g$; 
for AF $a-\lambda$ (violet);
for FM $g-f-\delta-\zeta$.
}
\end{figure}

(I) Evolution of the \textit{inhomogeneous} SF phases 
is given in Fig. 1.
At low fields,
due to the dominating role
of the exchange interactions favouring antiparallel ordering
of the magnetizations in adjacent layers,
some of the ``sublattices'' have to rotate 
against the applied field.
At sufficiently strong fields 
the sense of rotation for these ``sublattices'' 
is reversed (Fig. 1(b)). 
Near saturation, the SF phase  has only
positive projections of the magnetization 
on the direction of the magnetic field
which decreases towards the center 
similar to spin configurations described
in Ref. \cite{Nortermann92}.
There is a special field (independent of $N$)
where all inner sublattices have the same projection on the
field direction ($\theta_i= (-1)^{i+1}\,\theta_0$, 
$i=2, 3...N-1$) (Fig. 1(b)).
The parameters of this ``knot'' point are determined from 
the equations $H_{0}/J=(4-k)\cos\theta_0$, 
$\cos (2\theta_0) =k^{-1}-1/4 -\sqrt{1/16+k^{-2}}$,
$ \theta_1 + 3\,\theta_0 = \pi$, $k=K/J$.

(II) In the case of strong anisotropy, 
only collinear (Ising) states minimize the system energy. 
For Mills model, independently on $N$,
there are two discontinuous (``metamagnetic'') transitions: 
at $H_1 = J$ to the \textit{ferrimagnetic} phase
with flipped moment at both surfaces (FM) (Fig. 1(a)),
and between FM and ferromagnetic (F) phase at $H_2 = 2J$ (Fig. 2).

(III) A specific inhomogeneous  asymmetric 
\textit{canted} (C) phase (Fig. 1(a)) arises 
as a transitional low symmetry structure between
higher symmetry SF and FM phases. 
The transition FM $\to$ C is marked 
by the onset of noncollinearity, i.e. 
a deviation of $\mathbf{m}_i$ from the easy axis, 
and the transition SF $\to$ C breaks the mirror symmetry.

The calculated phase diagram with $N = 4$ 
in Fig. 2 includes all these phases 
and elucidates the corresponding magnetization processes 
for this Mills model.
The critical points $b$ and $f$ at $K_b \simeq 0.30$
and $K_f = \sqrt{2}$ for $N=4$ separate 
the phase diagram (Fig. 2) into three distinct regions.
In the \textit{low-anisotropy} region ($K<K_b$)
the first-order transition from AF to the 
inhomogeneous SF phase occurs 
at the critical line $a - b$, 
and a further second-order transition 
from SF into  F phase takes place at 
the higher field $H_f= (2+\sqrt{2})J-K$ 
(dashed line $g-f$ in Fig. 2).
In the \textit{high-anisotropy} region ($K>K_f$)
the above mentioned sequence of discontinuous
transitions AF $\to$ FM $\to$ F occurs.
In this region, different phases can exist 
as metastable states in extremely broad 
ranges of magnetic fields 
leading to severe hysteresis effects. 
Finally, in the intermediate region $K_b < K < K_f$
the magnetization processes have a complex
character including continuous and
discontinuous transitions into the C phase. 
For $N>4$ the region of the C-phase
is subdivided into smaller areas 
corresponding to canted asymmetric phases
separated by first-order critical lines
and an area of the reentrant SF phase (Fig. 3).
The number of these areas 
increases with increasing $N$. 
Here, the evolution of magnetic
states occurs as a cascade of
discontinuous transitions between different C-phases.

\begin{figure}
\includegraphics[width=9.25cm]{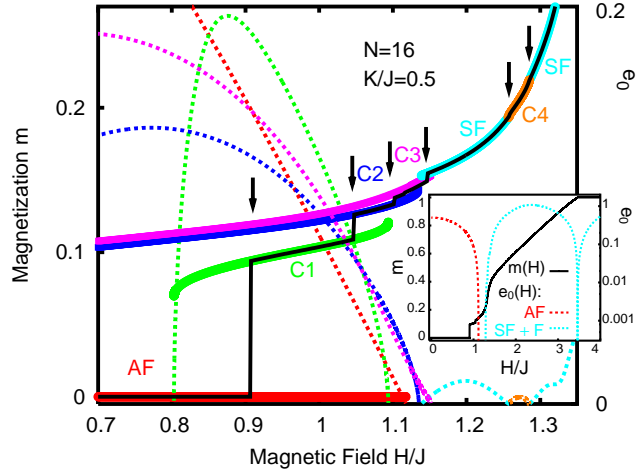}
\caption{
\label{MandEVvsHN16}
(color) 
Example of evolution of magnetization 
(continuous lines and left scales) 
and lowest eigenvalue of stability matrices 
$e_0$ (dotted, right scales)
for Mills model ($N=16$ and $K=0.5$). 
Black curves: magnetization of equilibrium states, 
color curves for various ``canted'' phases C1 $\dots $ C4 
and the (reentrant) spin-flop state ``SF''. 
Arrows mark phase transitions.
Inset gives full range of field H from 
antiferromagnetic (AF) to ferromagnetic (F) 
state (half-logarithmic plot for $e_0(H)$) -
details are magnified in main figure.
}
\end{figure}

Generally, the function (\ref{energy0}) can be considered
as the energy of a ``multi-sublattice'' antiferromagnet
with $N$ sublattices each represented by individual 
ferromagnetic layers. The phase diagram of such
an ``antiferromagnet'' in the space of the magnetic
parameters in the model (\ref{energy0}) may include
a number of new homogeneous and inhomogeneous phases
and additional phase transitions. 
In particular, for nonequal exchange constants 
there is a cascade of discontinuous transitions 
between different ferrimagnetic phases, 
and exchange anisotropy $K'_i$ may stabilize 
a \textit{twisted} phase \cite{InFight03}. 
Moreover, magnetic first-order transitions are generally 
accompanied by an involved reconstruction of multidomain structures 
and hysteresis \cite{Hubert98}, which will crucially determine 
the magnetic properties of experimental multilayer systems.
However, the basic features of the model (\ref{energy0})
are mainly imposed by cut exchange bonds and
are revealed from Mills model.
The phase diagram in Fig. 2 provides the backbone 
for the phase diagrams of the whole class of 
such nanostructures and is representative for
their magnetic states.

Our results show that
Mills model with finite $N$ 
owns only well-defined ``volume'' phases 
and transitions between them,
i.e. phases and transitions 
affecting the whole layer-stack. 
The model does not include solutions 
for surface-confined states, 
which were assumed to occur 
at a ``surface spin-flop field''
$H_{AF}= \sqrt{2JK+K^2}$ and to spread 
into the depth of the sample 
as the applied field increases up 
to the  ``bulk spin-flop field'' $H_{B}= \sqrt{4JK+K^2}$
\cite{Felcher02,Mills68}.
The critical field $H_{AF}$ determines 
the stability limit of the ``volume'' AF phase
(violet line $a-\lambda$ in Fig. 2), while
the field $H_{B}$
has no physical significance for the finite system.
Non-collinear inhomogeneous structures similar 
to those discussed here as SF phase have been observed 
in low anisotropic Fe/Cr superlattices \cite{Lauter02}.
The evolution of multidomain structures
accompanying spin-flop transitions
was investigated in  \cite{Nagy02}.
Inhomogeneous  asymmetric magnetic configurations 
found in Fe/Cr(211) superlattices with rather large
uniaxial anisotropy \cite{Wang94} and \cite{Felcher02}
are similar to C phases discussed in our paper.
The magnetization curve Fig. 3 for Mills
model with $N= 16$ and $K/J=0.5$ 
amends similar calculations (cf. Fig. 1 (a) in \cite{Wang94}).
In addition to the transition from AF into the C-phase,
the above described cascade of first-order transitions 
between different C-phases occurs.
A peculiarity of $m(H)$ 
interpreted as the bulk spin-flop field
(in Ref. \cite{Wang94} at $H$ = 1.49~kG $\widehat{=}H/J=1.49$)
does not correspond to any phase transition.

In conclusion, cut exchange bonds at the boundaries of 
antiferromagnetic superlattices
cause inhomogeneous, noncollinear, or canted  
magnetic configurations 
unknown in other types of magnetic nanostructures.
Experimental investigations (in particular 
on superlattices with small number of layers, 
$N$ = 4 and 6) should provide 
an interesting play-ground
to observe 
the rich variety of orientational effects 
predicted in this paper (Fig. 2).

\begin{acknowledgments}
A.\ N.\ B.\ thanks H.\ Eschrig for support and
hospitality at the IFW Dresden.
\end{acknowledgments}


\end{document}